\documentclass[aps,prd,floatfix,twocolumn,footinbib,altaffilletter]{revtex4-1}

\usepackage[colorlinks=true,citecolor=blue,linkcolor=blue]{hyperref}
\usepackage[utf8]{inputenc}
\usepackage{amsmath,amssymb}
\usepackage{epsfig}  
\usepackage{graphicx}               
\usepackage{url}
\usepackage{hyperref}
\usepackage{color}
\usepackage{placeins}
\usepackage[update,prepend]{epstopdf}
\usepackage{soul}

\usepackage{titlesec}
\titleformat{\section}{\normalfont\bfseries}{\thesection}{1em}{} 
\titlespacing*{\section}{0pt}{8pt}{8pt}

\def\thesection{\Roman{section}}

\begin{document}

\title{On the Charm Contribution to the Atmospheric Neutrino Flux}

\author{Francis Halzen}
\author{Logan Wille}
\email{lwille@icecube.wisc.edu}
\affiliation{Wisconsin IceCube Particle Astrophysics Center and Department of Physics, University of Wisconsin, Madison, WI 53706, USA} 

\begin{abstract}

We revisit the estimate of the charm particle contribution to the atmospheric neutrino flux that is expected to dominate at high energies because long-lived high-energy pions and kaons interact in the atmosphere before decaying into neutrinos. We focus on the production of forward charm particles which carry a large fraction of the momentum of the incident proton. In the case of strange particles, such a component is familiar from the abundant production of $K^+\Lambda$ pairs. These forward charm particles can dominate the high-energy atmospheric neutrino flux in underground experiments. Modern collider experiments have no coverage in the very large rapidity region where charm forward pair production dominates. Using archival accelerator data as well as IceCube measurements of atmospheric electron and muon neutrino fluxes, we obtain an upper limit on forward $\bar{D}^0\Lambda_c$ pair production and on the associated flux of high-energy atmospheric neutrinos. We conclude that the prompt flux may dominate the much-studied central component and represent a significant contribution to the TeV atmospheric neutrino flux. Importantly, it cannot accommodate the PeV flux of high-energy cosmic neutrinos, nor the excess of events observed by IceCube in the 30--200 TeV energy range indicating either structure in the flux of cosmic accelerators, or a presence of more than one component in the cosmic flux observed.
\end{abstract}

\maketitle

\section{Introduction}
\label{sec:intro}

The production of charm hadrons by cosmic rays interacting in the Earth's atmosphere \cite{Enberg:2008,Gauld:2015kvh,Gauld:2015yia,Bhattacharya:2015jpa,Garzelli:2015psa,Pasquali:1998ji, Gondolo:1995fq,Gaisser:2013rx,Gelmini:2000wm,Martin:2003bu} is the dominant background for the detection of cosmic neutrinos above an energy that depends on the charm cross section and on its dependence on Feynman $x_F$. Because of their short lifetime, charm hadrons decay promptly into neutrinos in contrast with relatively long-lived high-energy pions and kaons that interact and lose energy before decaying. Although prompt neutrinos may represent the dominant component of the atmospheric neutrino background for the identification of the cosmic neutrino flux at PeV energy, they have not yet been identified as such. IceCube observations \cite{JVS:2014} indicate that the neutrino flux is dominated by conventional atmospheric neutrinos at low energy and by cosmic neutrinos at high energy; charm neutrinos is expected to never dominate the measured spectrum. The issue is of great interest because a poor understanding of a potential charm neutrino background interferes with the precise characterization of the cosmic neutrino flux measured by IceCube.

We start by emphasizing that the production of charm in the atmosphere cannot accommodate the observed flux of high-energy cosmic neutrinos. We indeed know, independent of any theory, that the charm flux tracks the energy dependence of the cosmic ray flux incident on the atmosphere and that it is independent of zenith angle. A variety of analyses agree on the fact that there is no evidence for such a component in the IceCube data \cite{JVS:2014,Aartsen:2015zva}. While the flux above 200\,TeV can be accommodated by a power law with a spectral index $\gamma=2.07\pm0.13$ \cite{Schoenen:2015}, lowering the threshold revealed an excess of events in the 30--200\,TeV energy range \cite{JVS:2014}, raising the possibility the cosmic neutrino flux is not a single power or an additional charm background.

The production of charm particles has been extensively studied in the context of perturbative QCD \cite{Beenakker:1988bq,Ellis:1988sb,Nason:1987xz,Nason:1989zy}. These calculations often use a color dipole description of the target proton \cite{Arguelles:2015wba,Nikolaev:1995ty,Raufeisen:2002ka,Kopeliovich:2002yv} in order to mitigate the breakdown of the perturbative calculation associated with large $log(1/x)$ contributions, where $x=m_c / \sqrt{s}$. Here, $m_c$ is the charm quark mass and $s$ the center-of-mass energy of the colliding hadrons. At high energy, the charm quark is no longer a heavy quark whose mass controls the perturbative expansion. More importantly, these calculations only describe the central production of charm particles with a cross section that peaks at Feynman $x_F \sim 0$, providing an incomplete calculation. For strange particles, the central component of particle production is accompanied by a forward component where the incident proton transfers most of its energy to a $K^+ \Lambda$ pair with the same quantum numbers \cite{Edwards:1978mc, Abgrall:2015hmv}. It dominates strange particle production at large Feynman $x_F$. Forward charm production has been modeled with varying complexity, from intrinsic charm \cite{Brodsky:2015fna} and meson cloud description of the proton \cite{Steffens:1999hx} to the inclusion of QCD diagrams that promote one of the $c\bar{c}$ quarks in the proton to large Feynman $x_F$ when they hadronize with valence quarks in the incident proton \cite{Halzen:Diff}.

In this paper, we investigate the potential contribution of forward charm to atmospheric neutrino spectra. We do this by parameterizing the dependence of the charm cross section on Feynman $x_F$ and energy without reference to a specific model.  Also, the normalization is a free parameter. This parameterization is subsequently adjusted to accelerator and atmospheric neutrino data, which results in an upper limit on the forward charm contribution to the atmospheric neutrino flux at high energies. The forward component thus obtained contributes qualitatively at the same level as the central component to the total charm particle cross section, as is the case for strange particles. However, while it does potentially dominate the production of the highest energy atmospheric neutrinos in IceCube, we conclude that it cannot accommodate the flux of cosmic neutrinos that dominates the spectrum at the highest neutrino energies. In addition, this forward charm production is unable to accommodate the 30--200\,TeV excess over the best-fit power law seen in recent IceCube analyses that have lowered the threshold of the search for cosmic neutrinos \cite{JVS:2014}.

While we make no prediction for prompt neutrinos from forward charm, if produced at the level of the upper limit obtained here, the prompt spectrum could extend to higher energies than predicted by calculations that have neglected the forward component. While we conclude that the upper limit on the prompt neutrino flux is subdominant to the cosmic neutrino flux at all energies, it potentially represents a background, and it is therefore still important to characterize it.

The remainder of this paper is organized as follows: in Section \ref{sec:diff}, the parameterization of the differential cross section for the production of forward charm hadrons is introduced. In Section \ref{sec:flux}, we subsequently evaluate the upper limit on the flux of prompt neutrinos, and we confront it with the cosmic neutrino data in Section \ref{sec:conclu}.

\section{The Forward Charm Cross Section}
\label{sec:diff}

\begin{figure}
\includegraphics[width=0.5\textwidth]{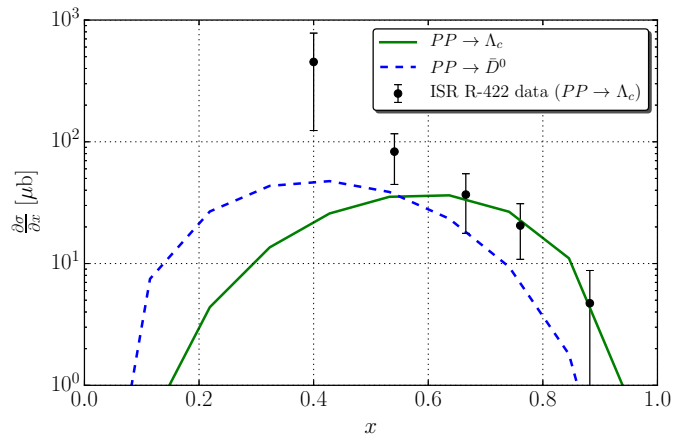}
\caption{The Feynman $x_F$ dependence for $\Lambda_c$ and $\bar{D}^0$ production using the parameterized cross section is compared with ISR data \cite{Bari:1991in} at $\sqrt{s} = 63$ GeV.}
\label{fig:shapes}
\end{figure}

To begin, we introduce a model-independent parameterization for forward charm production. It has the flexibility to adjust the energy and Feynman-x dependence independently: $\frac{d \sigma}{d x_F} = g(x_F)f(E_p)$. Specifically, the parameterization allows for changing the value at which the cross section peaks in $x_F$ while preserving the integrated cross section value, $\int_0^1 \, dx_F g(x_F) = \sigma$. The forward charm has been hypothesized to be produced by several processes each with a slightly different cross section peak. We initiate the calculation using a Feynman-x parameterization for forward $\Lambda_c$ and D production that peaks at large $x_F$ values, with a maximum at $x_F \sim \frac{2}{3}\,(\sim \frac{1}{3})$ for $\Lambda_c\,(D)$. These peak values are associated with the hadronization of charm quarks with the valence quarks in the incident proton. Without further adjustments this distribution matches the archival data on forward $\Lambda_c$ production from the CERN ISR $p\bar{p}$ collider \cite{Bari:1991in}. The differential cross sections are shown in Fig. \ref{fig:shapes} along with the ISR data that also fixes the normalization.

It has been argued that there is tension between different experimental results on the magnitude of forward charm production \cite{Hobbs:2013bia}. In the spirit of producing an upper limit, we use the ISR data, which measured the largest forward charm component \cite{Bari:1991in}.

For the energy dependence of the forward charm cross section we consider parameterizations bracketed by two extreme possibilities: the energy dependence of the total inelastic cross section $pp \rightarrow X$ and the inclusive charm cross section $pp \rightarrow c\bar{c} + X$ measured for centrally produced charm particles. We will refer to these as ``inelastic'' and $c\bar{c}$ dependence, respectively. In addition, we averaged the two as an illustration for an intermediate energy dependence.

\section{An Upper Limit on the Prompt Neutrino Flux}
\label{sec:flux}
To calculate the prompt neutrino spectrum from the decay of forward charm particles produced in the atmosphere, we have used the MCEq atmospheric interaction package \cite{Fedynitch:2015zma} in conjunction with a parameterization of the incident cosmic ray flux \cite{Gaisser:2012}. Observations indicate an increasing mass of the cosmic rays at the knee \cite{Apel:2011mi}. Heavier nuclei primaries shift the neutrinos from charm to lower energies, causing a break in the neutrino spectrum.

\begin{figure}
\includegraphics[width=0.5\textwidth]{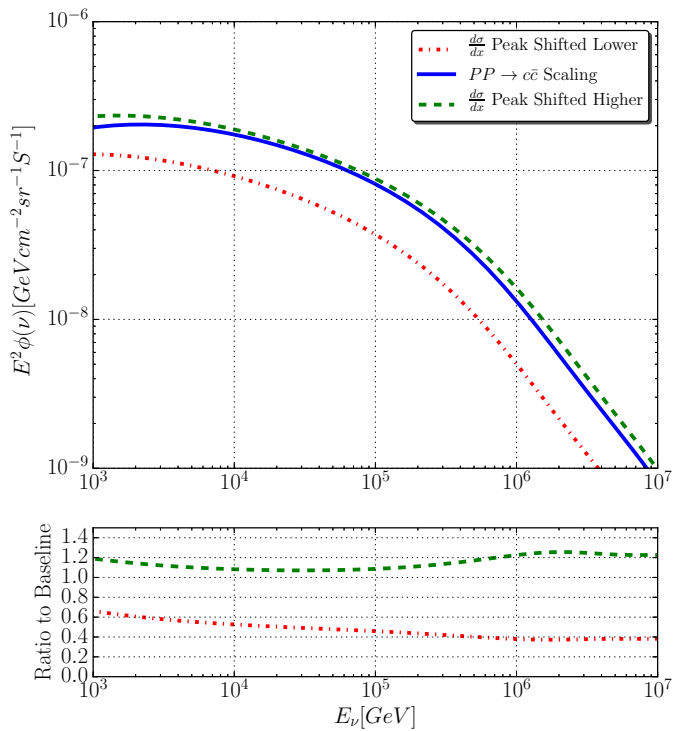}
\caption{The prompt neutrino spectrum from forward charm is shown using a baseline differential cross section and parameterizations with the $x_F$ maximum shifted up and down by 25\%. The ratio of the baseline and the two shifted cross sections is also shown. Note that the break in the spectrum occurs at different energies for the shifted cross sections.}
\label{fig:xfFlux}
\end{figure}

The result is shown in Fig. \ref{fig:xfFlux} assuming the $c\bar{c}$ energy dependence. The variation of the forward flux  on the detailed Feynman-x dependence of the cross section is illustrated by varying the position of its maximum. Shifting its value higher by 25\% has a small effect on the prompt neutrino flux that is already saturated by the initial parameterization. Lowering the peak value reduces the normalization without changing the spectrum in the region of interest beyond the break in energy associated with the ``knee'' in the cosmic ray spectrum.

\begin{figure}[t]
\includegraphics[width=0.5\textwidth]{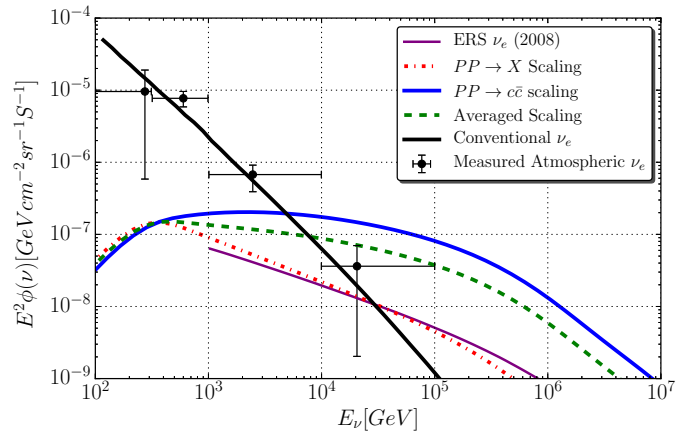}
\caption{The prompt electron neutrino spectrum from forward charm is shown for extreme assumptions of the energy dependence. Also shown is the result for an intermediate dependence that exceeds the measured flux \cite{Aartsen:2015xup} at the $1\,\sigma$ level at the highest energy of 20 TeV. An estimate of the contribution from centrally produced charm particles by Enberg. et al. (ERS) \cite{Enberg:2008} is shown for comparison.}
\label{fig:scFlux}
\end{figure}

Next we investigate the dependence of the prompt flux on the energy dependence of the cross section; see Fig. \ref{fig:scFlux}. The variations only affect the spectrum of the prompt neutrinos and not the normalization as all fluxes are equal at $\sim 3 \times 10^2$\,GeV. In this plot, we also compare the fluxes to the conventional neutrino flux from $\pi,K$ decays \cite{Honda:2006qj} and the flux of atmospheric electron neutrinos measured by IceCube \cite{Aartsen:2015xup}. The $c\bar{c}$ energy dependence exceeds the measured atmospheric neutrino flux but shows that the spectrum cannot mimic an $E^{-2}$ spectrum in the energy ranges of interest, $E_\nu > 10^4$ GeV. The ``averaged'' energy dependence does exceed the measured atmospheric neutrino flux at the $1\,\sigma$ level. In addition, each cross section shows a break in the spectrum $\sim 10^5$ GeV, reflecting the break in the cosmic ray spectrum at the knee.

\begin{figure*}
\includegraphics[width=0.95\textwidth]{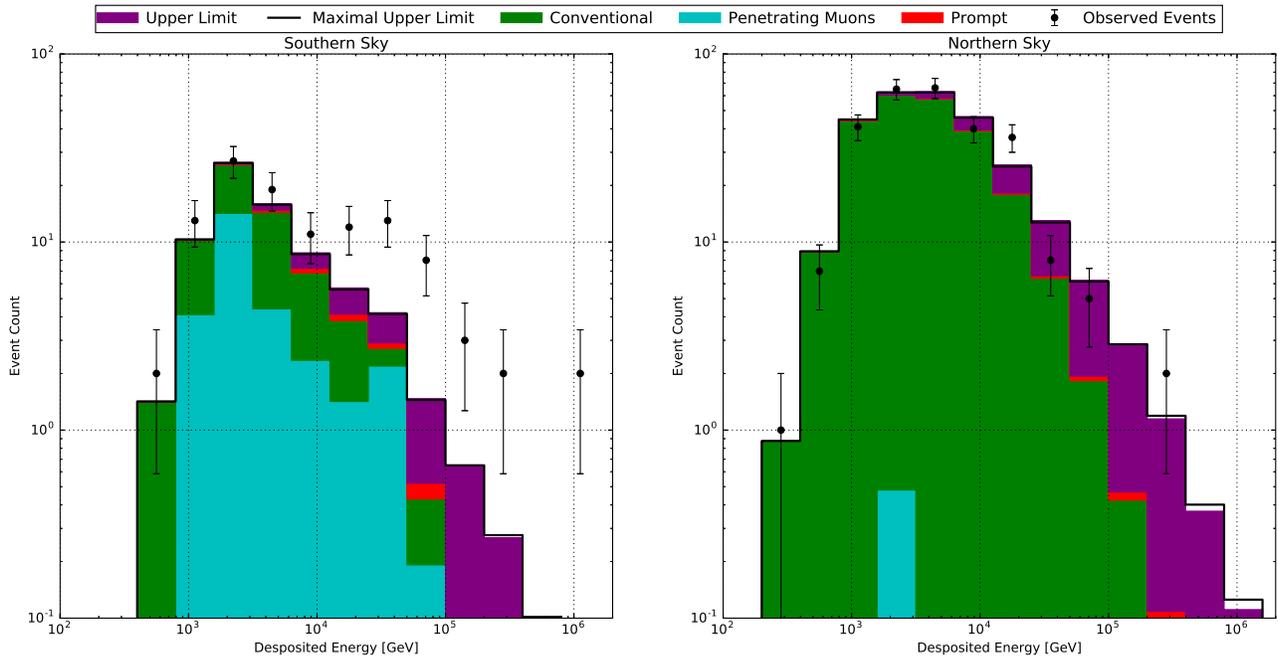}
\caption{The expected number of events in both the northern and southern sky for two years in IceCube using a veto-based detection scheme \cite{JVS:2014}. The upper limit fluxes include the self-veto effect as prescribed in \cite{PhysRevD.90.023009}. Neither upper limit can explain the high-energy events observed in IceCube.}
\label{fig:max}
\end{figure*}

It is clear that the parameterization with the ``averaged'' energy dependence represents an upper limit on the charm contribution at $1\,\sigma$. It is a conservative upper limit given that the flux in this energy region can be perfectly accommodated by the contributions from $\pi$ and K decays. To illustrate the strength of this upper limit, we show a second prompt flux with the maximal parameters, a steeper $c\bar{c}$ energy dependence with a shifted maximum in the $x_F$ distribution. This requires an adjustment of the normalization in order to not exceed the data. We now no longer match the ISR data. Both possibilities are confronted in Fig. \ref{fig:max} with the expected number of events in an IceCube starting event analysis (MESE) \cite{JVS:2014}. Although it does not qualitatively affect our conclusions, we have included the self-veto that requires the neutrino to not be accompanied by a detected atmospheric shower \cite{PhysRevD.90.023009, Schonert:2008is}.

Confronting the upper limit on prompt neutrino production to the observed IceCube events, we conclude that prompt neutrinos can possibly contribute to the flux in the 30--100 TeV range but not above 100 TeV, where neutrinos from cosmic origin dominate the data. Independently, the prompt flux simply traces the atmospheric cosmic ray spectrum and cannot accommodate the highest energy events observed in either analysis.

\section{Conclusions}
\label{sec:conclu}
We have used a parameterized cross section to model the forward component of charm production. It is expected to dominate the charm contribution to the high-energy atmospheric neutrino flux based on experience with strange particle production. We maximized its contribution to the atmospheric neutrino flux by varying both its Feynman-$x_F$ and energy dependence without exceeding data from collider and high-energy atmospheric neutrino experiments. We subsequently calculated the upper limit of the flux of prompt neutrinos from the decay of charmed particles in IceCube, which is dominated by the forward component of the flux. We found that the prompt neutrino flux from forward charm may represent a background to the cosmic neutrino flux but cannot explain the high-energy events observed by IceCube at energies above 100\,TeV.

\section{Acknowledgments}

Discussion with collaborators inside and outside the IceCube Collaboration, too many to be listed, have greatly shaped this presentation. We thank A. Fedynitch for assistance in using MCEq and Paolo Lipari for useful discussion. This research was supported in part by the U.S. National Science Foundation under Grants No.~ANT-0937462 and PHY-1306958 and by the University of Wisconsin Research Committee with funds granted by the Wisconsin Alumni Research Foundation.

\bibliographystyle{apsrev}
\bibliography{Letter}

\end{document}